\documentclass[usenatbib]{mnras}
\usepackage{amssymb,amsmath}
\usepackage{graphicx}
\usepackage{xcolor,natbib}
\usepackage{multirow}
\usepackage[T1]{fontenc}
\usepackage{ae,aecompl}
\usepackage{newtxtext, newtxmath}
\usepackage{float}
\usepackage{subfigure}
\usepackage{longtable}
\usepackage{rotating}
\usepackage{enumitem}
\usepackage{longtable,listings}
\usepackage[flushleft]{threeparttable}
\usepackage{parcolumns}
\usepackage{lineno}

\def\o3{[O~{\sc iii}]\ }

\title[Test R-L reltion for NLRs of AGN]{To test $R_{NLRs}~-~L_{O3}$ relation for narrow 
emission line regions of AGN through low redshift Type-2 AGN in SDSS}
\author[Zhang]
{Xue-Guang Zhang\thanks{Corresponding author Email:
            \href{mailto:xgzhang@gxu.edu.cn}{xgzhang@gxu.edu.cn}} \\
	    Guangxi Key Laboratory for Relativistic Astrophysics, School of Physical Science 
	    and Technology, GuangXi University, Nanning, 530004, P. R. China}

\date{}

\begin{document}
\pagerange{\pageref{firstpage}--\pageref{lastpage}} \pubyear{2024}
\maketitle
\label{firstpage}

\begin{abstract}
	Sizes of narrow emission line regions (NLRs) of AGN could be estimated by \o3 line 
luminosity $L_{O3}$ through the known $R_{NLRs}-L_{O3}$ empirical relations. Unfortunately, 
it is not convenient to test the $R_{NLRs}-L_{O3}$ empirical relations through structure 
properties of spatially resolved NLRs of large samples of AGN. In this manuscript, a method 
is proposed to test the $R_{NLRs}-L_{O3}^{\sim0.25}$ empirical relations for AGN NLRs through 
SDSS Type-2 AGN having few orientation effects on NLRs sizes expected by AGN unified model, after 
considering sizes $R_{fib}$ of SDSS fiber covered regions. Comparing $R_{fib}$ and $R_{NLRs}$ 
estimated by $L_{O3}$, Type-2 AGN with $R_{fib}>R_{NLRs}$ (Sample-II) and with $R_{fib}<R_{NLRs}$ 
(Sample-I) should have different physical properties of NLRs. Accepted electron density gradients 
in AGN NLRs, statistically higher electron densities (traced by lower flux ratio $R_{S2}$ of 
[S~{\sc ii}]$\lambda6717$\AA~ to [S~{\sc ii}]$\lambda6731$\AA) could be expected for the 
Type-2 AGN in the Sample-I. Then, through the collected 1062 SDSS Type-2 AGN in the Sample-I 
and 3658 SDSS Type-2 AGN in the Sample-II, statistically lower $R_{S2}$ for the Type-2 AGN in 
the Sample-I can be confirmed with confidence level higher than 5$\sigma$, even after considering 
necessary effects. Therefore, the results in this manuscript can provide strong clues to support 
that the reported $R_{NLRs}~\propto~L_{O3}^{0.25}$ empirical relation is preferred to estimate 
NLRs sizes of SDSS AGN through SDSS fiber spectroscopic results, and also to support the commonly 
expected electron density gradients in AGN NLRs. 
\end{abstract}

\begin{keywords}
galaxies:active - galaxies:nuclei - galaxies:emission lines - galaxies:Seyfert
\end{keywords}

\section{Introduction}

	Narrow emission line regions (NLRs) at kilo-parsec (kpc) scales to central black holes 
(BHs) are fundamental structures of active galactic nuclei (AGN) \citep{nl93, ns04, sf09, 
sh11, dh15, gg17, td18, mf19, mc23}. So large spatial distance to central BHs and extended 
structures, NLRs in nearby AGN with high luminous \o3 emissions can be well spatially resolved, 
leading to well determined sizes of AGN NLRs ($R_{NLRs}$, distance between NLRs to central BHs). 
Then, the known $R_{NLRs}-L_{O3}$ empirical relation is proposed to estimate $R_{NLRs}$ 
through well measured \o3 line luminosity $L_{O3}$, 
\begin{equation}
\log(\frac{R_{NLRs}}{\rm pc})\propto~0.25\log(\frac{L_{O3}}{\rm 10^{42}erg/s}) 
\end{equation}.
The equation above is firstly reported in \citet{lz13} through Gemini Integral Field Unit 
observations of spatially resolved [O~{\sc iii}] emission regions in a sample of low redshift 
($z\sim0.5$) radio quiet Type-2 quasars, and then well discussed in \citet{hh13, hh14} for an 
upper limit for $R_{NLRs}$ in high luminous quasars. \citet{sg18} have reported 
the similar dependence of $R_{NLRs}$ on $L_{O3}^{0.22-0.33}$ through a new broad-band imaging 
technique to reconstruct the spatial structures of the [O~{\sc iii}] emission regions of 300 
obscured AGN at redshifts 0.1-0.7. \citet{hs22} have shown the similar dependence (but with 
large scatters) of $R_{NLRs}$ on $L_{O3}^{0.25-0.29}$ through wide-field optical 
integral-field unit spectroscopy for 41 luminous unobscured AGN with $z<0.06$ in the Close 
AGN Reference Survey. More recently, we \citet{zh22} have tested the 
$R_{NLRs}-L_{O3}^{\sim0.25}$ empirical relation through a sample of AGN with double-peaked 
broad Balmer emission lines.

	Meanwhile, besides the reported $R_{NLRs}-L_{O3}$ empirical relation with slope about 
0.25\ in \citet{lz13, sg18, hs22}, there are $R_{NLRs}-L_{O3}$ relations with different slopes 
found in the literature. \citet{sc03, fk18} have reported one $R_{NLRs}-L_{O3}$ relation with 
slope about 0.4-0.5, through Hubble Space Telescope (HST) observations of [O~{\sc iii}] 
emission regions in samples of Seyfert 2 galaxies and/or Type-2 quasars. As discussed in 
\citet{lz13}, different $R_{NLRs}-L_{O3}$ relation in \citet{sc03, fk18} from the one in 
\citet{lz13} could be due to HST observations being 1-2 magnitudes shallower than the Gemini 
Integral Field Unit observations and long-slit spectroscopy applied in \citet{lz13}. 
More recently, \citet{cs19} have reported $R_{NLRs}-L_{O3}^{\sim0.42}$ through 
spatially resolved structures of extended narrow line regions of a large sample of nearby AGN 
from the Mapping Nearby Galaxies at Apache Point Observatory survey.

	Furthermore, besides the discussed $R_{NLRs}-L_{O3}$ empirical relations through 
observations, more recent review on theoretically expected $R_{NLRs}-L_{O3}$ relations can be 
found in \citet{dz18} through modelling AGN NLRs as a collection of clouds in pressure 
equilibrium with the ionizing radiation. The $R_{NLRs}-L_{O3}$ empirical relations reported 
in \citet{lz13} and in \citet{fk18} can be both theoretically explained by different model 
parameters in \citet{dz18}.  
	
	Until now, only a few tens of AGN have their NLRs spatially resolved leading to well 
estimated $R_{NLRs}$, to test the predicted $R_{NLRs}-L_{O3}$ empirical relations through 
NLRs properties in a large sample of AGN should be very interesting and necessary. 
Unfortunately, it is not convenient to spatially resolve NLRs in samples of AGN. Therefore, 
in this manuscript, an indirect method is proposed to test $R_{NLRs}-L_{O3}$ empirical 
relations through a large sample of low redshift Type-2 AGN in SDSS, after considering natural 
applications of 3\arcsec~ fiber diameters for SDSS spectroscopic results and properties of 
AGN unified model expected projected space size of NLRs in Type-2 AGN, which is the main 
objective of this manuscript.

	The known AGN unified model has been well discussed in \citet{an93, bm12, mb12, Oh15, 
nh15, ma16, aa17, bb18, bn19, kw21, zh22, zh23} to well explain different observational 
phenomena between broad line AGN and narrow line AGN (Type-2 AGN). The commonly accepted AGN 
unified model indicates few orientation effects on estimated $R_{NLRs}$ in Type-2 AGN. 
Therefore, a large sample of low redshift Type-2 AGN collected from Sloan Digital Sky Survey 
(SDSS) are mainly considered in the manuscript. Moreover, considering fixed fiber diameters 
for SDSS spectra, fiber expected projected space sizes can be applied to test different 
physical properties of NLRs in Type-2 AGN, as well described in Section 2.

	In this manuscript, section 2 presents our main hypotheses. Section 3 shows our main 
results and necessary discussions. Section 4 gives our main summary and conclusions. And in 
the manuscript, the cosmological parameters of $H_{0}~=~70{\rm km\cdot s}^{-1}{\rm Mpc}^{-1}$, 
$\Omega_{\Lambda}~=~0.7$ and $\Omega_{m}~=~0.3$ have been adopted.

\section{Main Hypotheses}

	Due to fixed fiber diameter 3\arcsec~ for SDSS spectra\footnote{As the collected 
Type-2 AGN, discussed in Section 3.1, are all from SDSS not from eBOSS with fixed fiber 
diameter 2\arcsec. Therefore, there are no further discussions on Type-2 AGN collected from 
eBOSS in this manuscript.}, sizes of \o3 emissions ($R_{NLRs}$, distance between NLRs and 
central BHs) in SDSS spectrum of a Type-2 AGN can be discussed by the following two cases. 
First, if \o3 emissions of a Type-2 AGN were totally covered by the SDSS fiber, intrinsic 
$R_{NLRs}$ (with few effects of orientation effects) should be smaller than 
$R_{fib}=1.5$\arcsec, meanwhile the measured \o3 line luminosity was the intrinsic value. 
Second, if \o3 emissions of a Type-2 AGN were partly covered by the SDSS fiber, intrinsic 
$R_{NLRs}$ should be larger than $R_{fib}=1.5$\arcsec, meanwhile, the measured \o3 line 
luminosity should be smaller than the intrinsic value.

	Comparing estimated $R_{NLRs}$ by observed \o3 line luminosity and fiber expected 
$R_{fib}$, the collected SDSS Type-2 AGN can be divided into two samples, one sample 
(sample-I) of Type-2 AGN with $R_{NLRs}~>~R_{fib}$ and the other sample (sample-II) of Type-2 
AGN with $R_{NLRs}~<~R_{fib}$. Then, considering dependence of spatially resolved electron 
density on distance of emission regions to central BHs as well shown and discussed in 
\citet{kg18, kb19}, i.e. apparent electron density gradients in AGN NLRs, SDSS Type-2 AGN 
in the Sample-1 should have statistically higher electron densities determined through 
spectroscopic features than SDSS Type-2 AGN in the sample-II, due to NLRs with lower electron 
densities not covered by the SDSS fibers for spectra of Type-2 AGN in the sample-I. Certainly, 
if there were no apparent electron density gradients in AGN NLRs, i.e. constant 
electron density distributions in AGN NLRs, some further properties on NLRs should be further 
proposed in this manuscript. Therefore, it is interesting to check spectroscopic properties 
of SDSS Type-2 AGN in the Sample-I and in the Sample-II.

	Furthermore, it is very convenient to estimate electron density in AGN NLRs, through 
emission line flux ratio $R_{S2}$ of [S~{\sc ii}]$\lambda6716$\AA~ to 
[S~{\sc ii}]$\lambda6731$\AA~ as well discussed in \citet{sa16, kn17, kb19, kn19, fm20, kn21, 
rr21, dv22, zh23}. Therefore, once large samples of SDSS Type-2 AGN are created, properties 
of electron density in AGN NLRs can be easily determined, which will provide further clues 
not only to (or not to) support the reported $R_{NLRs}-L_{O3}$ empirical relations with 
different slopes in the literature but also to (or not to) support electron density gradients 
in AGN NLRs.  

	Finally, if accepted electron density gradients in AGN NLRs, considering $R_{S2}$ 
to trace electron densities in AGN NLRs (lower $R_{S2}$ indicating higher electron densities), 
we would have
\begin{equation}
R_{S2}(Sample-I)~<~R_{S2}(Sample-II)
\end{equation},
which is the basic point applied to test $R_{NLRs}-L_{O3}$ empirical relations in the manuscript.

\section{Main Results and Necessary Discussions}

\subsection{Type-2 AGN in main sample}

	As well described in \url{https://www.sdss.org/dr16/spectro/catalogs/}, each 
spectroscopic object has a classification in SDSS. Simple criteria can be well applied to 
conveniently collect all the low redshift pipeline classified Type-2 AGN from SDSS main 
narrow emission line galaxies in data release 16 (DR16) \citep{ap20}, through the SDSS 
provided SQL (Structured Query Language) Search tool 
(\url{http://skyserver.sdss.org/dr16/en/tools/search/sql.aspx}) by the following query
\begin{lstlisting}
SELECT plate, fiberid, mjd, z, G.sii_6717_flux, 
   G.sii_6731_flux, G.oiii_5007_flux, 
   G.nii_6584_flux, G.h_alpha_flux,
   G.h_beta_flux, G.oiii_4363_flux, 
   G.oiii_4363_flux_err
FROM SpecObjall as S JOIN GalSpecLine as G 
   on S.specobjid = G.specobjid
WHERE
   class='galaxy' and subclass = 'AGN' and 
   z < 0.3 and zwarning = 0 
   and G.sii_6717_flux_err > 0 and 
   G.sii_6717_flux > 5*G.sii_6717_flux_err 
   and G.sii_6731_flux_err > 0 and
   G.sii_6731_flux > 5*G.sii_6731_flux_err
   and G.oiii_5007_flux_err > 0 and 
   G.oiii_5007_flux > 5*G.oiii_5007_flux_err 
   and G.nii_6584_flux_err > 0 and 
   G.nii_6584_flux > 5*G.nii_6584_flux_err 
   and G.h_alpha_flux_err > 0 and 
   G.h_alpha_flux > 5*G.h_alpha_flux_err
   and G.h_beta_flux_err > 0 and
   G.h_beta_flux > 5*G.h_beta_flux_err
\end{lstlisting}
The SQL query above can lead 11803 narrow emission line galaxies (pipeline classified 
Type-2 AGN) to be collected into our main sample.

	In the query above, 'SpecObjall' is the SDSS pipeline provided database 
(\url{http://skyserver.sdss.org/dr16/en/help/docs/tabledesc.aspx}) including basic properties 
of spectroscopic features of emission line galaxies in SDSS DR16, 'GalSpecLine' is the 
database including measured line parameters of emission lines provided by the MPA-JHU group 
as well described in \url{https://www.sdss.org/dr16/spectro/galaxy_mpajhu/} and in \citet{bc04, 
ka03a, th04}. In the query above, class='galaxy' and subclass='AGN' mean SDSS spectrum can 
be well identified with a galaxy template and the galaxy has detectable emission lines that 
are consistent with being a Seyfert-2 or a LINER classified by BPT diagram \citep{bpt, kb01, 
ka03a, kb06, kb19, zh20}. In the query above, z < 0.3 and zwarning = 0 are applied to 
confirm [S~{\sc ii}] doublet covered by SDSS spectra with reliable pipeline determined 
redshift. In the query above, G.oiii\_5007\_flux\_err > 0 and G.oiii\_5007\_flux > 
5*G.oiii\_5007\_flux\_err and  G.sii\_6717\_flux\_err > 0 and G.sii\_6717\_flux > 
5*G.sii\_6717\_flux\_err and G.sii\_6731\_flux\_err > 0 and G.sii\_6731\_flux > 
5*G.sii\_6731\_flux\_err and G.nii\_6584\_flux\_err > 0 and G.nii\_6584\_flux > 
5*G.nii\_6584\_flux\_err and G.h\_alpha\_flux\_err > 0 and G.h\_alpha\_flux > 
5*G.h\_alpha\_flux\_err and G.h\_beta\_flux\_err > 0 and G.h\_beta\_flux > 
5*G.h\_beta\_flux\_err are applied to confirm reliable measurements of \o3, [S~{\sc ii}], 
[N~{\sc ii}] and narrow Balmer emission line fluxes which will be applied in this manuscript, 
considering measured line fluxes at least 5 times larger than their uncertainties.

	Before proceeding further, one point is noted. As what we have recently checked and 
shown in \citet{zh23}, the 'GalSpecLine' provided measured fluxes of narrow emission lines 
in Type-2 AGN are reliable enough, and there are no further discussions in this manuscript 
on effects of measurements of emission line fluxes on reliability of following statistical 
results.

\begin{figure}
\centering\includegraphics[width = 8cm,height=13cm]{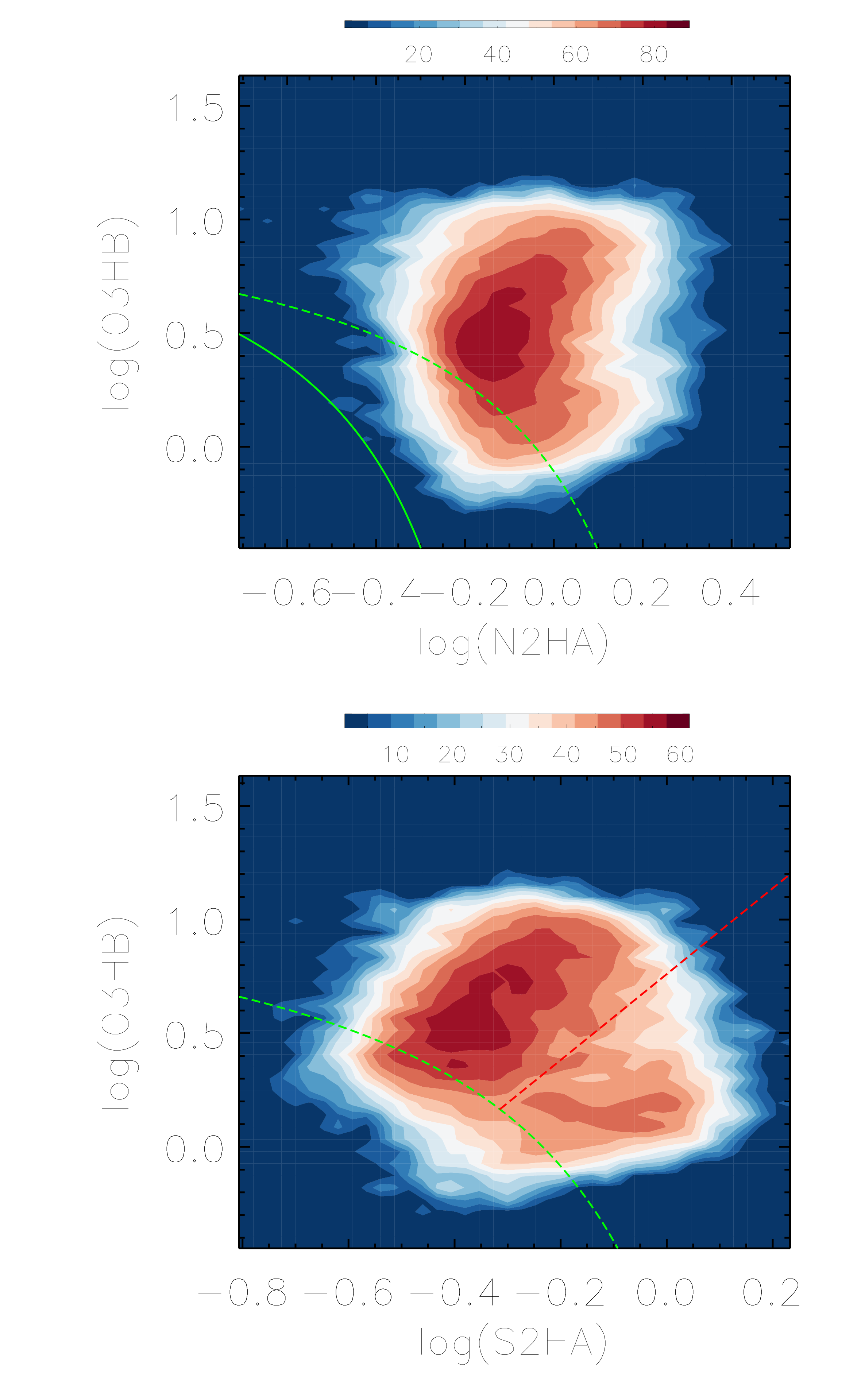}
\caption{BPT diagrams in contour of O3HB versus N2HA (top panel) and of O3HB versus S2HA 
(bottom panel) of the collected 11803 objects. Contour levels in different colors are 
related to number densities shown in color bar in top region of each panel. Dashed green 
line in each panel shows the extremely dividing line between AGN and HII galaxies reported 
in \citet{kb06}. Solid green line in top panel shows the dividing line between HII galaxies 
and composite galaxies reported in \citet{ka03}. Dashed red line in bottom panel shows the 
dividing line between AGN and LINERs reported in \citet{kb06}.}
\label{bpt}
\end{figure}

	Then, Fig.~\ref{bpt} shows BPT diagrams of flux ratio of \o3 to narrow H$\beta$ 
(O3HB) versus flux ratio of [N~{\sc ii}] to narrow H$\alpha$ (N2HA) and of O3HB versus 
flux ratio of [S~{\sc ii}] to narrow H$\alpha$ (S2HA) of the collected 11803 Type-2 AGN in 
our main sample. Due to SDSS pipeline provided AGN classifications (objects lying above 
solid green line in top panel of Fig.~\ref{bpt}) only through flux ratios of O3HB and N2HA, 
part of collected objects in the main sample are lying into HII regions (below dashed green 
line in bottom panel of Fig.~\ref{bpt}) in the BPT diagram of S2HA versus O3HB. Furthermore, 
as different physical mechanisms discussed in the literature \citep{ht80, ds96, ft92, eh10, 
csm11, mm17}, there are no confirmed conclusions on LINERs as genuine AGN. Therefore, in 
the manuscript, LINERs are not considered. The dividing line (dashed red line in bottom panel 
of Fig.~\ref{bpt}) in \citet{kb06} is applied to identify LINERs in the BPT diagram of S2HA 
versus O3HB. In order to ignore probable effects from starforming and LINERs, we only consider 
the 7355 objects classified as AGN both in the BPT diagram of N2HA versus O3HB and in the BPT 
diagram of S2HA versus O3HB, i.e., the objects lying not only above the dashed green line in 
top panel of Fig.~\ref{bpt} but also above the dashed red line in bottom panel of Fig.~\ref{bpt}.

\begin{figure}
\centering\includegraphics[width = 8cm,height=5.5cm]{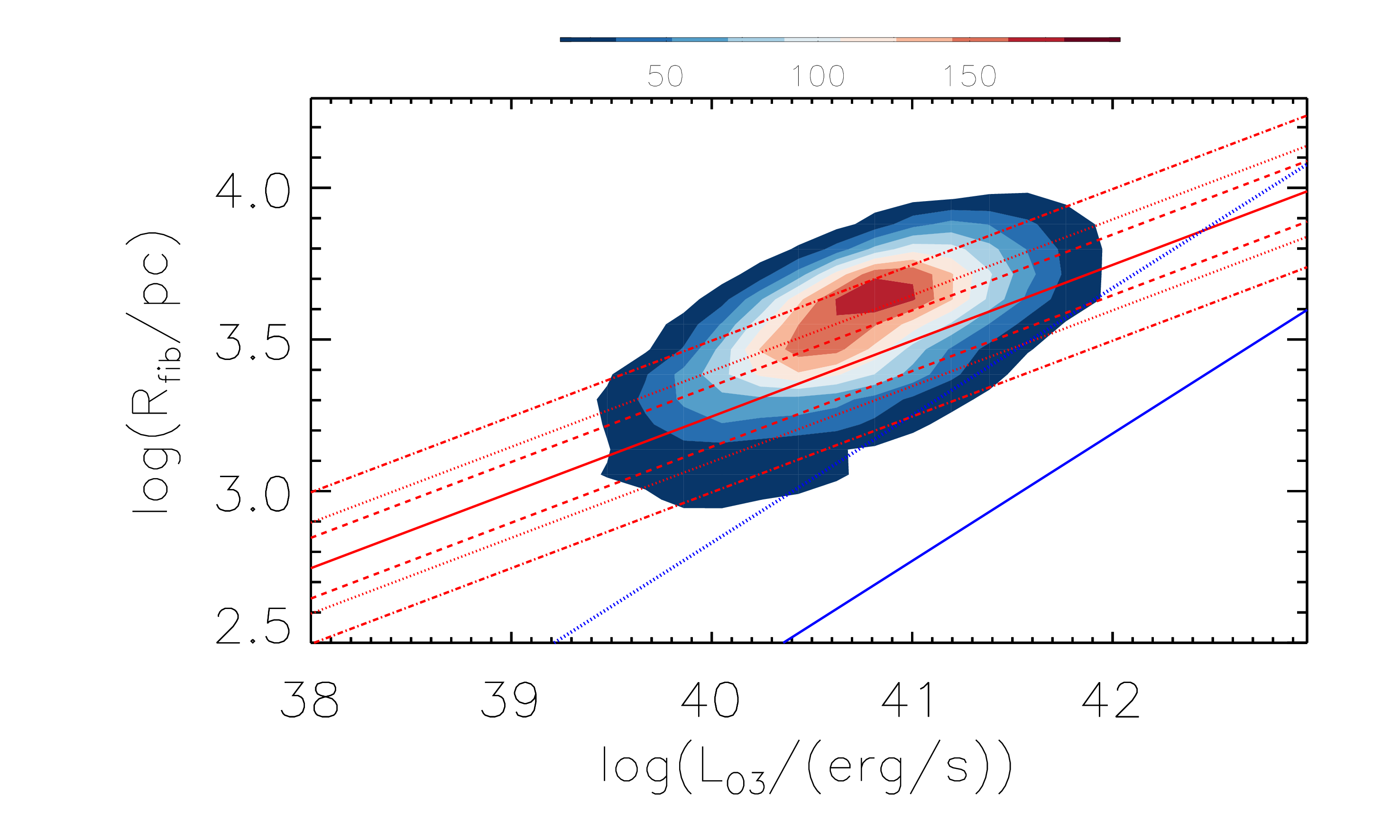}
\caption{Correlation in contour between SDSS fiber radii $R_{fib}$ in units of pc and observed 
$L_{O3}$ of the 7355 Type-2 AGN. Contour levels in different colors are related to number 
densities shown in color bar in top region. Solid red line, dashed red lines, dotted red 
lines and dot-dashed red lines show the $R_{NLRs}-L_{O3}$ relation reported in \citet{lz13} 
and corresponding 2RMS, 3RMS and 5RMS scatters. Solid blue line shows the reported 
$R_{NLRs}-L_{O3}$ relation in \citet{fk18}. Dotted blue line shows the reported $R_{NLRs}-L_{O3}$ 
relation in \citet{cs19}.}
\label{rl}
\end{figure}

	Then, Fig.~\ref{rl} shows dependence of $R_{fib}$ (in units of pc, calculated by fiber 
radius and corresponding redshift) on observed \o3 line luminosity of the collected 7355 
Type-2 AGN, and also the $R_{NLRs}-L_{O3}^{0.25}$ empirical relation in \citet{lz13}. 
Interestingly, part of Type-2 AGN have $R_{fib}$ clearly larger than $R_{NLRs}-L_{O3}$ empirical 
relation expected values, but part of Type-2 AGN have $R_{fib}$ apparently smaller than 
$R_{NLRs}-L_{O3}$ empirical relation expected values. As discussed in Section 2, Type-2 AGN 
above and below the $R_{NLRs}-L_{O3}$ empirical relation in Fig.~\ref{rl} should have 
expected different electron densities in their NLRs. Further discussions on 
$R_{NLRs}-L_{O3}^{0.4-0.5}$ will be given in Section 3. 

\begin{figure}
\centering\includegraphics[width = 8cm,height=6cm]{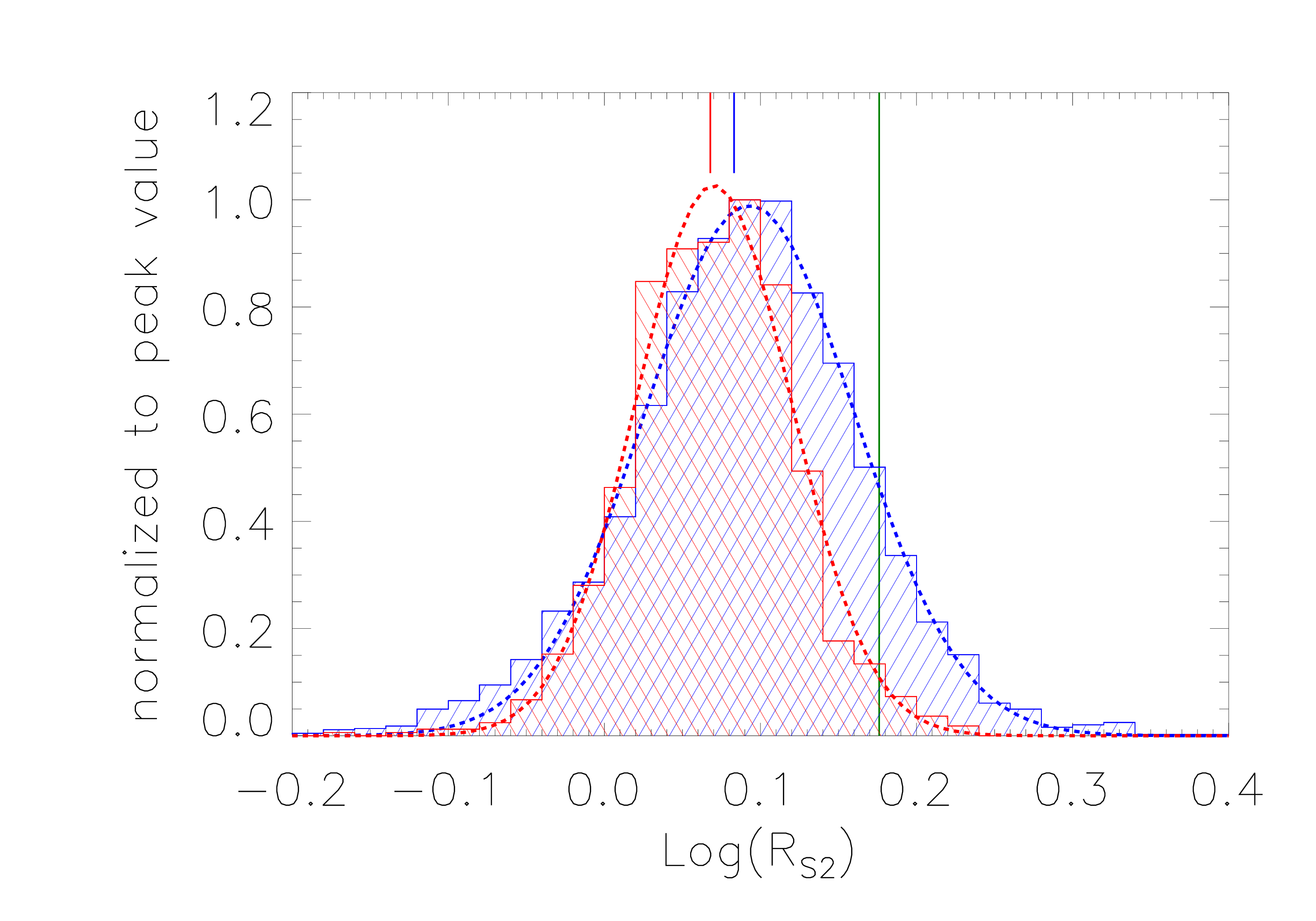}
\caption{Distributions of $\log(R_{S2})$ of the Type-2 AGN in the Sample-II (histogram filled 
by blue lines) and in the Sample-I (histogram filled by red lines). Thick dashed line shows 
Gaussian-like descriptions to the histogram in the same color. Vertical dark green line 
marks the position of $R_{S2}=1.5$. Vertical thick lines in blue and in red in top region mark 
positions of median values of $\log(R_{S2})$ of the Type-2 AGN with $R_{S2}<1.5$ in 
the Sample-II and in the Sample-I. 
}
\label{ne}
\end{figure}

\begin{figure}
\centering\includegraphics[width = 8cm,height=6cm]{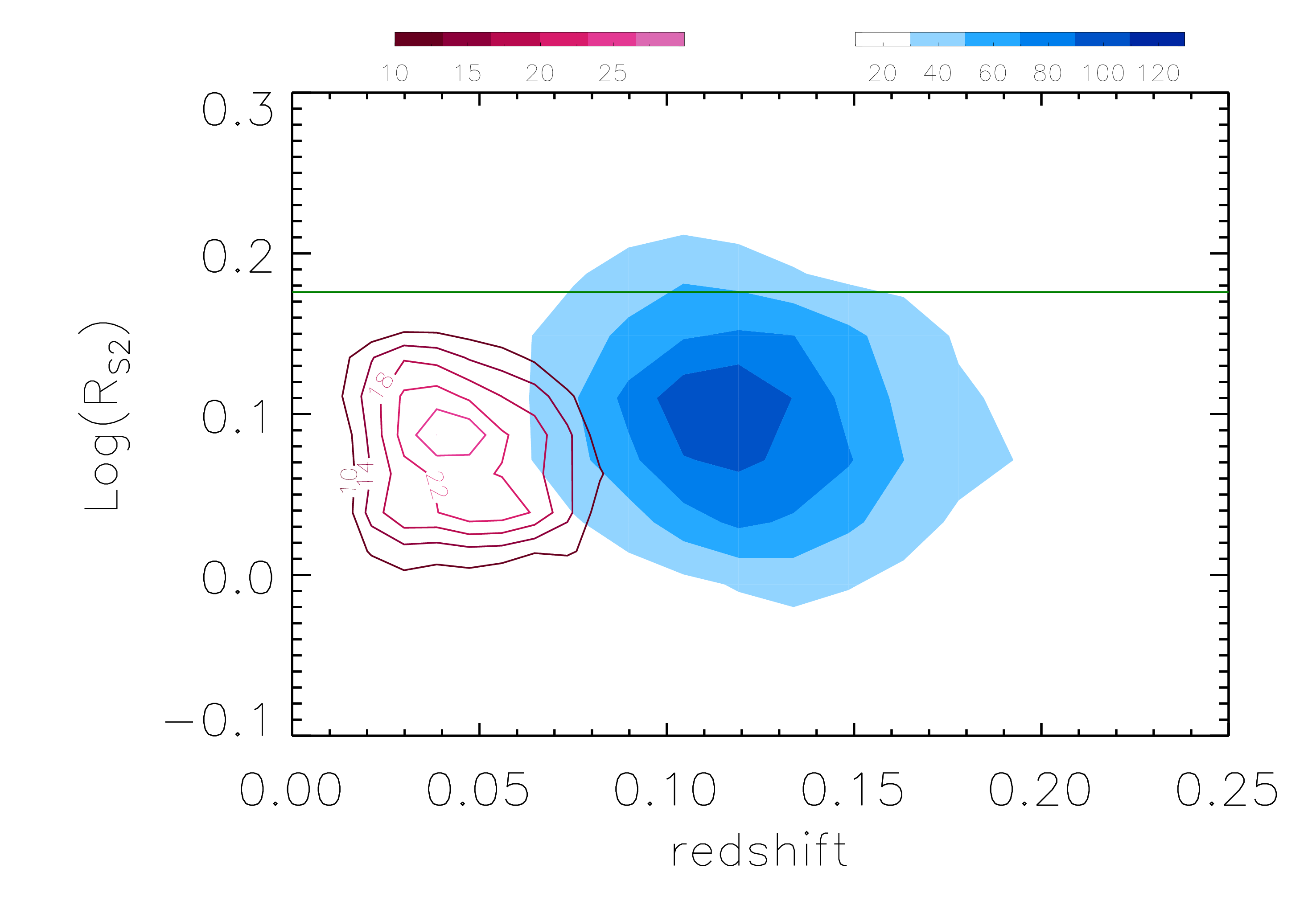}
\caption{On the dependence of $\log(R_{S2})$ on redshift for the Type-2 AGN in the Sample-II 
(contour filled by bluish colors) and in the Sample-I (contour levels labelled and shown in reddish 
colors). Contour levels in different colors are related to number densities shown in the color 
bars in the top region. Horizontal dark green line marks the position of $R_{S2}=1.5$.}
\label{zd}
\end{figure}

\begin{figure}
\centering\includegraphics[width = 8cm,height=10cm]{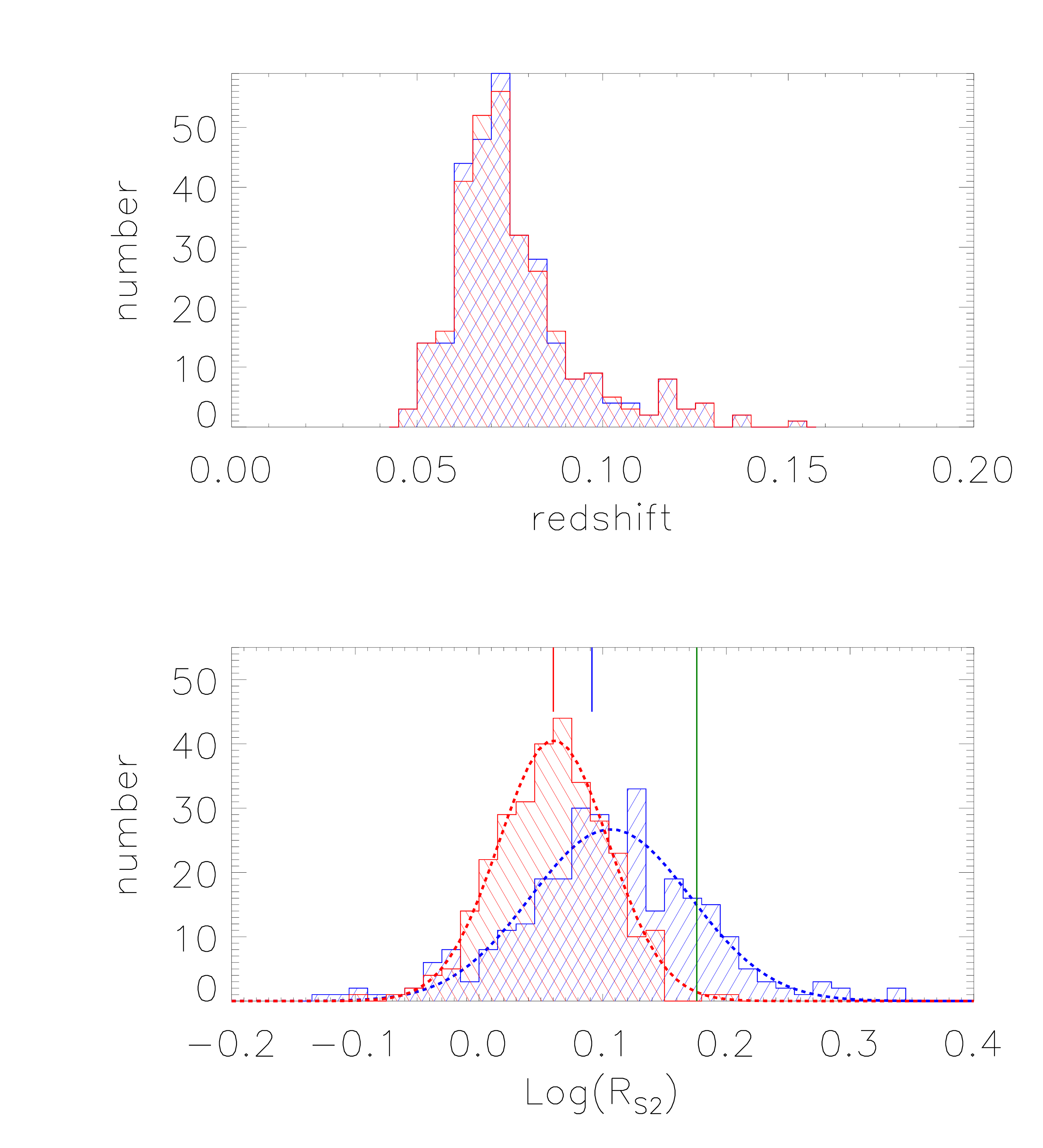}
\caption{Distributions of redshift (in top panel) and $\log(R_{S2})$ (in bottom panel) of 
the Type-2 AGN in the subsample-II (histogram filled by blue lines) and in the subsample-I 
(histogram filled by red lines). In bottom panel, thick dashed line shows Gaussian-like 
descriptions to the histogram in the same color, vertical dark green line marks the 
position of $R_{S2}=1.5$, vertical thick lines in blue and in red in top region mark positions 
of the median values of $\log(R_{S2})$ of the Type-2 AGN with $R_{S2}<1.5$ in the 
subsample-II and in the subsample-I.
}
\label{szd}
\end{figure}

\subsection{Main Results and Discussions}

	Based on Fig.~\ref{rl} and considering the RMS scatter about 0.05 of the $R_{NLRs}-L_{O3}$ 
empirical relation in \citet{lz13}, two samples of Type-2 AGN are created, one sample (Sample-II) 
of Type-2 AGN includes the 3658 Type-2 AGN lying above the $R_{NLRs}-L_{O3}$ empirical relation plus 
2RMS scatter and the other sample (Sample-I) of Type-2 AGN includes the 1062 Type-2 AGN lying below 
the $R_{NLRs}-L_{O3}$ empirical relation minus 2RMS scatter. Here, applications of $R_{NLRs}-L_{O3}$ 
relation plus/minus 2RMS scatter can not only lead as many Type-2 AGN as possible to be collected 
into following Sample-I and Sample-II, but also lead to few effects of scatters of the $R_{NLRs}-L_{O3}$ 
empirical relation on our following statistical results. And applications of $R_{NLRs}-L_{O3}$ relation 
plus/minus 3RMS scatters and 5RMS scatters will be discussed before the end of the subsection in this 
manuscript.

	Then, Fig.~\ref{ne} shows $R_{S2}$ distributions of the Type-2 AGN in the Sample-I and in 
the Sample-II. Meanwhile, as noted in \citet{kn17, kn19} that when $R_{S2}$ is applied 
to trace electron density, $R_{S2}$ should be effectively limited to the range from 0.4 to 1.5, i.e., 
$-0.4~<~\log(R_{S2})~<~0.18$. Therefore, objects with $\log(R_{S2})~>~0.18$ should be not considered 
to trace electron density. The median values of $\log(R_{S2})$ (with $R_{S2}<1.5$) are 
about $(8.31\pm0.13)\times10^{-2}$ and $(6.79\pm0.15)\times10^{-2}$ for the Type-2 AGN in the Sample-II 
and in the Sample-I, respectively. Uncertainties of the median values are estimated by the 
bootstrap method within 1000 loops, as what we have recently done in \citet{zh23}. For each loop, about 
half of the data points of $\log(R_{S2})$ ($R_{S2}<1.5$) of the Type-2 AGN in the Sample-I or in the 
Sample-II are randomly collected to create a new sample, leading to a re-measured median value. After 
1000 loops, based on distribution of the 1000 re-measured median values, half width at half maximum of 
the distribution is accepted as the uncertainty of the median value. It is clear that there are quite 
different median values of $\log(R_{S2})$ (with $R_{S2}<1.5$) for the Type-2 AGN in the Sample-II and 
in the Sample-I. Meanwhile, through the Wilcoxon Rank-Sum Test, the Type-2 AGN in the 
subsample-II and in the subsample-I have the same median values of $\log(R_{S2})$ (with $R_{S2}<1.5$) 
with significance level smaller than $10^{-17}$. Moreover, the lower median value of $\log(R_{S2})$ 
(with $R_{S2}<1.5$) in the Type-2 AGN in the Sample-I is consistent with what we can expected (as 
discussed in Section 2) for Type-2 AGN in Sample-I having their NLRs partly covered by SDSS fibers.

	Before proceeding further, the following seventh points should be discussed to confirm different 
$\log(R_{S2})$ (with $R_{S2}<1.5$) of the Type-2 AGN in the Sample-I and in the Sample-II.

	First, there are apparently different redshift distributions of the Type-2 AGN in the Sample-II 
and in the Sample-I, as shown in Fig.~\ref{zd} with mean redshift about 0.13 and 0.05. 
Therefore, it is necessary to discuss effects of different redshift distributions on comparing 
$\log(R_{S2})$ (with $R_{S2}<1.5$) shown in Fig.~\ref{ne} of the Type-2 AGN in the Sample-II and in 
the Sample-I. Then, the following two methods are applied. On the one hand, the dependence 
of $\log(R_{S2})$ on redshift shown in Fig.~\ref{zd} is firstly checked. The Spearman Rank correlation 
coefficients are about -0.12 ($P_{null}\sim8\times10^{-14}$) and -0.13 ($P_{null}\sim1\times10^{-5}$) 
for the Type-2 AGN in the Sample-II and in the Sample-I, respectively. Not different dependence of 
$\log(R_{S2})$ on redshift indicates few effects of different redshift distributions on the statistically 
different $\log(R_{S2})$ for the Type-2 AGN in the Sample-I and in the Sample-II. On the other hand, 
among the Type-2 AGN in the Sample-II and in the Sample-I, two subsamples, one subsample (subsample-II) 
including 301 Type-2 AGN from the Sample-II and the other subsample (subsample-I) including 301 Type-2 
AGN from the Sample-I, can be simply created to have the same redshift distributions with significance 
levels higher than 99.91\% through the two-sided Kolmogorov-Smirnov statistic technique, similar 
as what we have recently done in \citet{zh23} through finding the minimum parameter distance. The same 
redshift distributions of Type-2 AGN in the subsamples are shown in top panel of Fig.~\ref{szd}. Meanwhile, 
bottom panel of Fig.~\ref{szd} shows $\log(R_{S2})$ distributions of the Type-2 AGN in the two subsamples. 
The median values and the corresponding bootstrap method determined uncertainties of 
$\log(R_{S2})$ (with $R_{S2}<1.5$) are about $(9.13\pm0.24)\times10^{-2}$ and $(6.01\pm0.21)\times10^{-2}$ 
for the Type-2 AGN in the subsample-II and in the subsample-I, respectively. And through the 
Wilcoxon Rank-Sum Test, the Type-2 AGN in the subsample-II and in the subsample-I have the same median 
values of $\log(R_{S2})$ (with $R_{S2}<1.5$) with significance level smaller than $10^{-17}$. Therefore, 
there are no dependence of $\log(R_{S2})$ on redshift, indicating few effects of different redshift 
distributions on different $\log(R_{S2})$ for the Type-2 AGN in the Sample-II and in the Sample-I.

	Second, due to the discussed results above mainly based on collected Type-2 AGN in the Sample-I 
and the Sample-II through considering 2RMS scatters of the $R_{NLRs}-L_{O3}$ empirical relation in 
\citet{lz13}, it is necessary to check whether $k\times$RMS ($k$ larger than 2) scatters applied can 
lead to different results. Here, 3RMS (5RMS) scatters are applied to build a new Sample-I including 736 
(363) Type-2 AGN and a new Sample-II including 2746 (1135) Type-2 AGN. Considering 3RMS (5RMS) scatters 
can lead median values and the corresponding bootstrap method determined uncertainties of 
$\log(R_{S2})$ (with $R_{S2}<1.5$) to be about $(7.88\pm0.17)\times10^{-2}$ and $(6.32\pm0.21)\times10^{-2}$ 
($(7.08\pm0.24)\times10^{-2}$ and $(5.93\pm0.23)\times10^{-2}$) for the Type-2 AGN in the new Sample-II 
and in the new Sample-I, respectively. And considering 3RMS (5RMS) scatters, the Wilcoxon Rank-Sum 
Test can be applied to confirm significantly different median values of $\log(R_{S2})$ with significance 
level about $10^{-17}$ ($10^{-8}$) (confidence level higher than 5$\sigma$) of the Type-2 AGN in the new 
Sample-I and in the new Sample-II. Therefore, different criteria on RMS scatters have few effects on the 
different median $\log(R_{S2})$ of the Type-2 AGN in the Sample-I and in the Sample-II. 

\begin{figure}
\centering\includegraphics[width = 8cm,height=5cm]{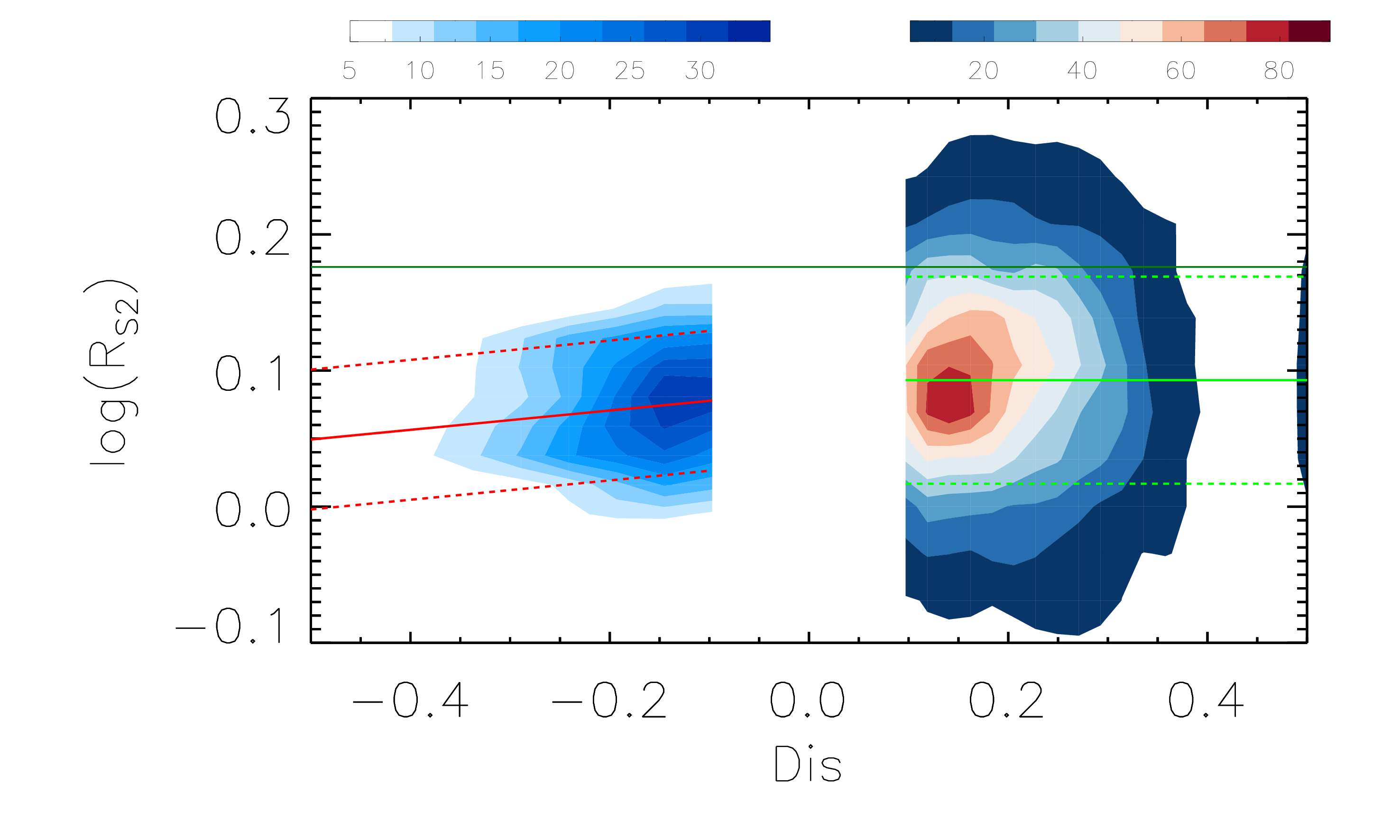}
\caption{Dependence of $R_{S2}$ on distance $Dis$ of objects to the $R_{NLRs}-L$ empirical relation 
in \citet{lz13}. Contour in left region shows the results for the 1062 Type-2 AGN in the Sample-I, 
and contour in right region shows the results for the 3658 Type-2 AGN in the Sample-II. Solid red line 
and dashed red lines show the linearly dependence and the corresponding 1RMS scatter of $\log(R_{S2})$ 
on $Dis$ for the Type-2 AGN in the Sample-I. Solid green and dashed green lines mark the position of 
the median value of $Dis$ and the corresponding 1RMS scatter of the Type-2 AGN in the 
Sample-II. Horizon dark green line marks the position of $R_{S2}=1.5$.}
\label{dep}
\end{figure}

\begin{figure*}
\centering\includegraphics[width = 18cm,height=5cm]{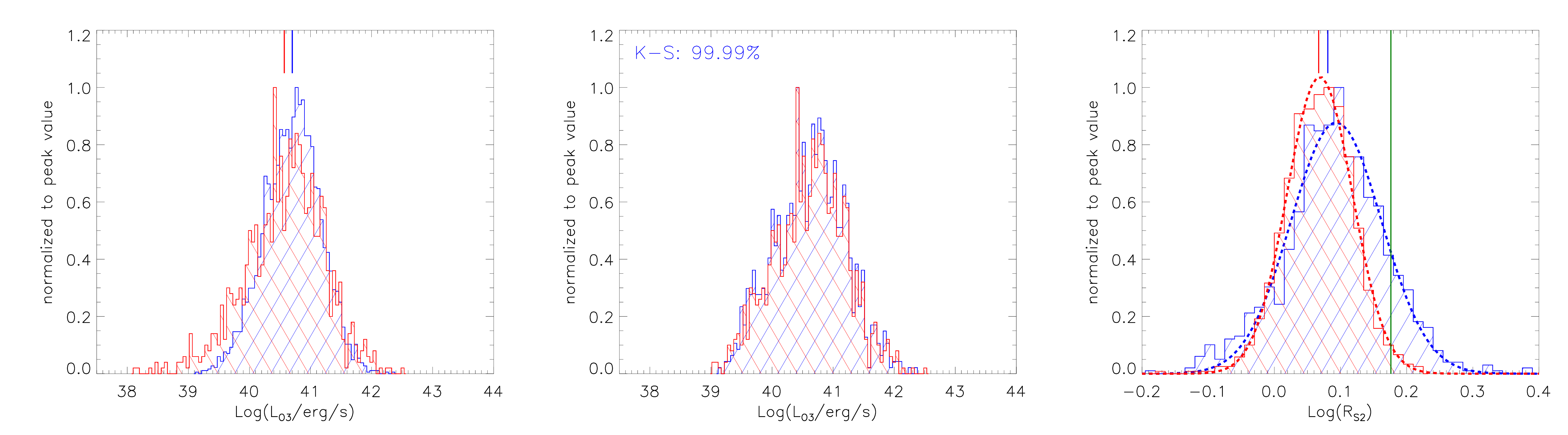}
\caption{Left panel shows the $L_{O3}$ distributions of the Type-2 AGN in the Sample-II (histogram filled 
by blue lines) and in the Sample-I (histogram filled by red lines), respectively. Vertical blue line and 
red line in top region of left panel mark the mean values of $\log(L_{O3})$ of the Type-2 AGN in the 
Sample-II and in the Sample-I, respectively. Middle panel shows the $L_{O3}$ distributions of the Type-2 
AGN in the new subsamples which have the same $L_{O3}$ distributions. Right panel shows the $\log(R_{S2})$ 
distributions of the Type-2 AGN in the new subsamples which have the same $L_{O3}$ distributions. In the 
right panel, line styles have the same meanings as those in Figure~\ref{ne}.}
\label{d2p}
\end{figure*}

\begin{figure*}
\centering\includegraphics[width = 18cm,height=4cm]{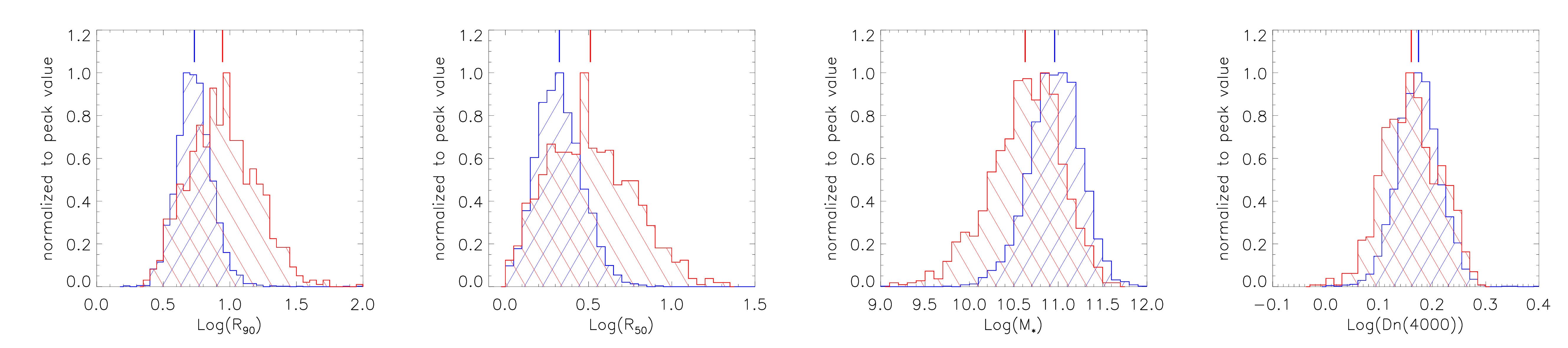}
\caption{Distributions of the $R_{90}$, $R_{50}$, $M_{*}$ and $Dn(4000)$ of the 
Type-2 AGN in the Sample-II (histogram filled by blue lines) and in the Sample-I (histogram 
filled by red lines). Vertical blue line and red line in top region of each panel mark the 
mean values of the parameter of the Type-2 AGN in the Sample-II and in the Sample-I, respectively.}
\label{n1ew}
\end{figure*}

\begin{figure*}
\centering\includegraphics[width = 18cm,height=8cm]{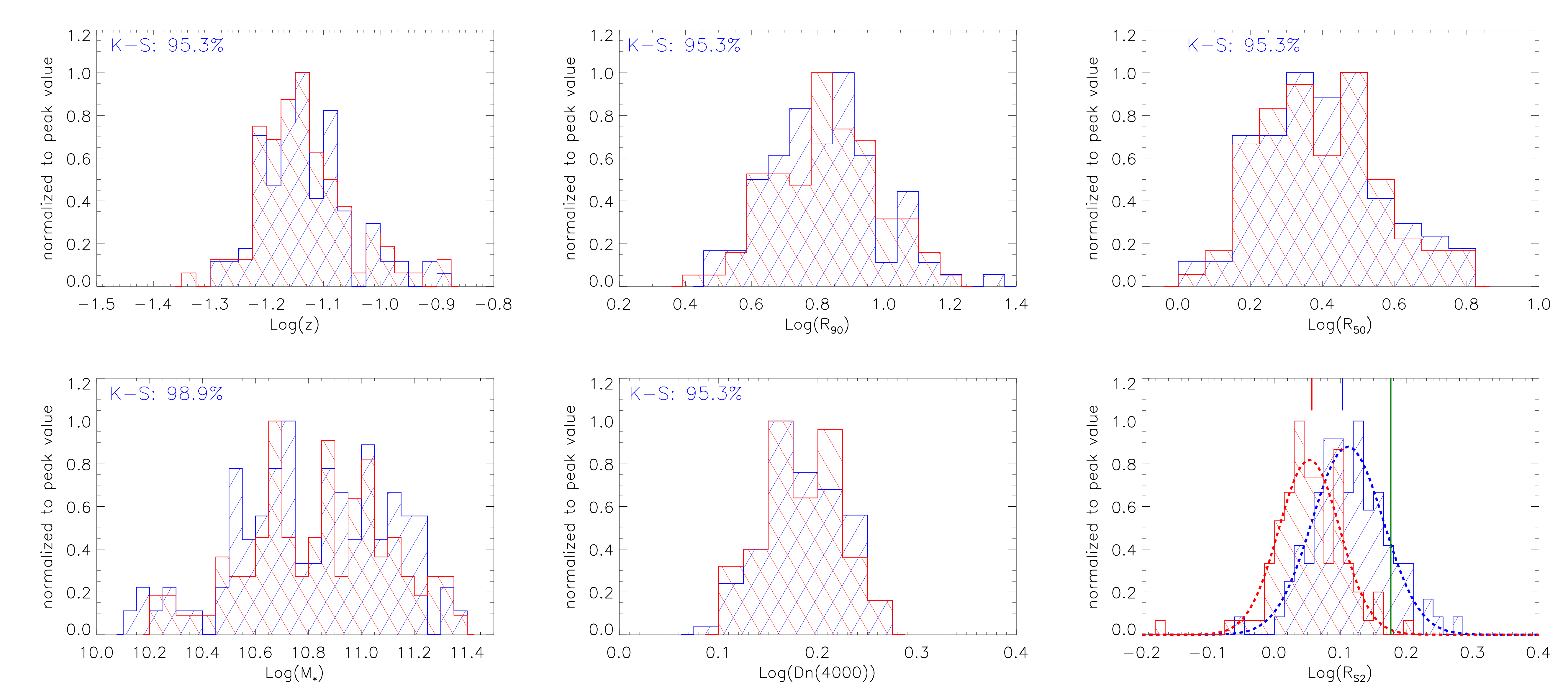}
\caption{The first five panels show the distributions of redshift, $R_{90}$, $R_{50}$, $M_{*}$ 
and $Dn(4000)$ of the 96 Type-2 AGN in the new subsample-II (histogram 
filled by blue lines) and of the 96 Type-2 AGN in the new subsample-I (histogram filled by 
red lines). The significance level is marked in top region of each panel for the two 
distributions being the same through the two-sided Kolmogorov-Smirnov statistic technique. 
The bottom right panel shows the $R_{S2}$ distributions of  the 96 Type-2 AGN in the new 
subsample-II (histogram filled by blue lines) and of the 96 Type-2 AGN in the new subsample-I 
(histogram filled by red lines). In the bottom right panel, line styles have the same meanings 
as those in Figure~\ref{ne}.}
\label{n2ew}
\end{figure*}

	Third, dependence of $R_{S2}$ on distance $Dis$ of objects to the $R_{NLRs}-L_{O3}$ empirical 
relation in \citet{lz13} is checked for the 3658 Type-2 AGN in the Sample-II and the 1062 Type-2 AGN 
in the Sample-I, and shown in Fig.~\ref{dep}. Here, Type-2 AGN in Sample-II above the $R_{NLRs}-L_{O3}$ 
empirical relation have positive values of $Dis$, Type-2 AGN in Sample-I below the $R_{NLRs}-L$ empirical 
relation have negative values of $Dis$. For the Type-2 AGN in the Sample-II, there is no dependence of 
$R_{S2}$ on distance $Dis$, with calculated Spearman Rank correlation coefficient about 0.003 
($P_{null}\sim0.84$). Therefore, if NLRs totally covered by SDSS fibers, no dependence of $R_{S2}$ on distance 
$Dis$ can be determined. Meanwhile, for the Type-2 AGN in the Sample-I, there is acceptable dependence 
of $R_{S2}$ on distance $Dis$: $\log(R_{S2})~\propto~0.07~\times~Dis$, with calculated Spearman Rank 
correlation coefficient about 0.16 ($P_{null}\sim3\times10^{-7}$). Therefore, the accepted dependence 
for the Type-2 AGN in the Sample-I but no dependence for the Type-2 AGN in the Sample-II can provide 
clues to support the basic point that whether NLRs totally covered by SDSS fibers can lead to statistically 
different properties of NLRs.

	Fourth, as well discussed and shown in \citet{of06, kn19, fm20, zh23}, there are apparent effects 
of electron temperature on electron density estimated by $R_{S2}$, it is necessary to check whether are 
there different electron temperature properties for the Type-2 AGN in the Sample-I and in the Sample-II. 
Electron temperatures $T_e$ in AGN NLRs can be well traced by the emission line flux ratio $O_{32}$ of total 
[O~{\sc iii}]$\lambda4959,5007$\AA~ to [O~{\sc iii}]4363\AA. Unfortunately, as shown in \citet{zh23}, not 
like [S~{\sc ii}] doublet commonly apparent in Type-2 AGN, only a small part of Type-2 AGN have apparent 
[O~{\sc iii}]$\lambda4363$\AA~ emissions. Based on the 'GalSpecLine' provided emission line fluxes and 
uncertainties of [O~{\sc iii}]4363\AA~ of the Type-2 AGN in the Sample-I and the Sample-II, there are only 
7 of 3658 Type-2 AGN in the Sample-II and 289 of 1062 Type-2 AGN in Sample-I having their 
[O~{\sc iii}]$\lambda4363$\AA~ emission fluxes at least five times larger than their uncertainties. 
Therefore, it is hard to provide statistical results on properties of $O_{32}$ of the Type-2 AGN in the 
Sample-I and in the Sample-II. However, simply qualitative discussions can be given as follows. Based on 
the electron temperature $T_e$ estimated through $O_{32}$
\begin{equation}
O_{32}~\propto~\exp(\frac{3.29\times10^4}{T_e}) 
\end{equation},
quite weak [O~{\sc iii}]$\lambda4363$\AA~ emissions (not apparent in SDSS spectra) in the Type-2 AGN in 
the Sample-II could indicate larger $O_{32}$, leading to lower electron temperature in NLRs for the Type-2 
AGN in Sample-II than in Sample-I: $T_e(Sample-II)~<~T_e(sample-I)$. Meanwhile, accepted the dependence 
of electron density $n_e$ on $T_e$ and $R_{S2}$
\begin{equation}
\frac{n_e}{\rm cm^{3}}\times(\frac{10^4K}{T_e})^{0.5}~\approxeq~
	f(R_{S2})=\frac{627.1~\times~R_{S2}~-~909.17}{0.4315~-~R_{S2}}
\end{equation},
considering that the Type-2 AGN in the Sample-II having smaller $f(R_{S2})$ (larger $R_{S2}$) and 
smaller $T_e$ than the Type-2 AGN in the Sample-I, it is apparent that very larger electron density 
$n_e$ could be expected in NLRs of the Type-2 AGN in the Sample-I after considering effects of electron 
temperature on estimations of $n_e$ through $R_{S2}$, well consistent with expected result that higher 
electron densities in NLRs of the Type-2 AGN in the Sample-I than in the Sample-II.

	Fifth, similar as done to check effects of different redshift distributions, effects of different 
$L_{O3}$ distributions are checked as follows. Left panel of Fig.~\ref{d2p} shows the 
$\log(L_{O3}/{\rm erg/s})$ distributions with the mean value about 40.72 of the 3658 Type-2 AGN in the 
Sample-II and with the mean value about 40.57 of the 1062 Type-2 AGN in the Sample-I, respectively. And 
through the Students t-statistic technique, the Type-2 AGN in the Sample-II and in the Sample-I have the 
same mean values with significance level only about $8\times10^{-17}$, indicating it is necessary to check 
effects of different $L_{O3}$ distributions. Among the Type-2 AGN in the Sample-II and in the Sample-I, two 
subsamples, one subsample (subsample-II) including 1033 Type-2 AGN from the Sample-II and the other 
subsample (subsample-I) including 1033 Type-2 AGN from the Sample-I, can be simply created to have the 
same $\log(L_{O3})$ distributions with significance levels higher than 99.99\% through the two-sided 
Kolmogorov-Smirnov statistic technique. The same $\log(L_{O3})$ distributions of Type-2 AGN in the new 
subsamples are shown in middle panel of Fig.~\ref{d2p}. Meanwhile, right panel of Fig.~\ref{d2p} shows 
$\log(R_{S2})$ distributions of the Type-2 AGN in the two new subsamples. The median values 
and the corresponding bootstrap method determined uncertainties of $\log(R_{S2})$ (with $R_{S2}<1.5$) are 
about $(8.08\pm0.14)\times10^{-2}$ and $(6.71\pm0.15)\times10^{-2}$ for the Type-2 AGN in the new 
subsample-II and in the new subsample-I, respectively. And through the Wilcoxon Rank-Sum Test, 
the Type-2 AGN in the new subsample-II and in the new subsample-I have the same median values of 
$\log(R_{S2})$ (with $R_{S2}<1.5$) with significance level only about $10^{-15}$, to re-confirm higher 
$R_{S2}$ for the Type-2 AGN above the $R_{NLRs}-L_{O3}$ relation. Therefore, there are few effects of 
different $\log(L_{O3})$ distributions on different $\log(R_{S2})$ for the Type-2 AGN in the Sample-II 
and in the Sample-I.

\begin{figure*}
\centering\includegraphics[width = 18cm,height=5cm]{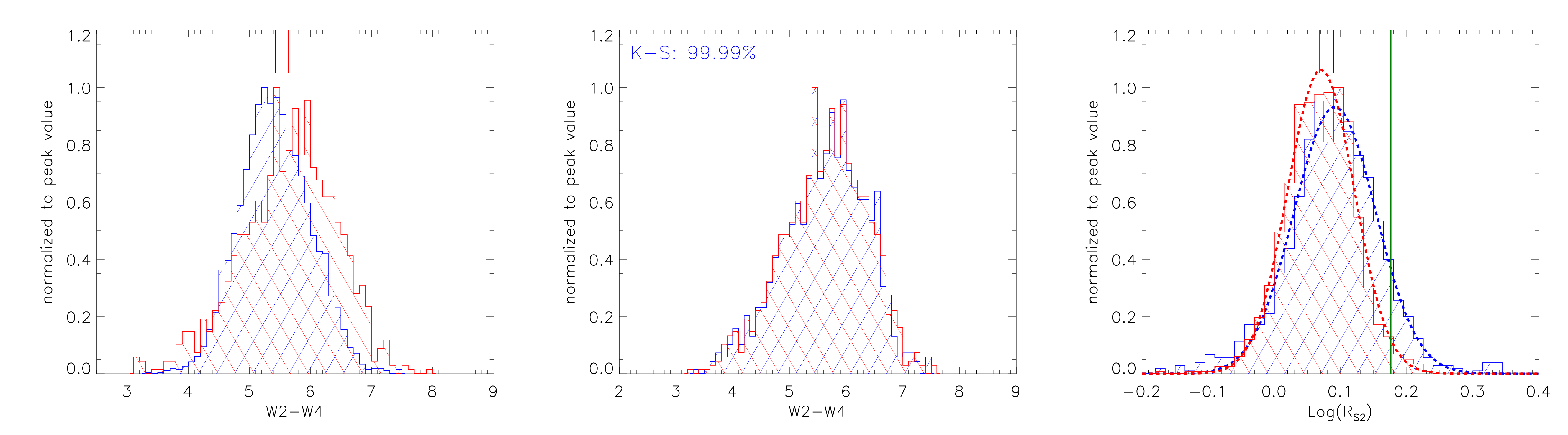}
\caption{Left panel shows the W2-W4 distributions of the Type-2 AGN in the Sample-II (histogram filled 
by blue lines) and in the Sample-I (histogram filled by red lines), respectively. Vertical blue line 
and red line in top region of left panel mark the mean values of W2-W4 of the Type-2 AGN in the Sample-II 
and in the Sample-I, respectively. Middle panel shows the W2-W4 distributions of the Type-2 AGN in the 
new subsamples which have the same W2-W4 distributions. Right panel shows the $\log(R_{S2})$ distributions 
of the Type-2 AGN in the new subsamples which have the same W2-W4 distributions. In the right panel, line 
styles have the same meanings as those in Figure~\ref{ne}.}
\label{wise}
\end{figure*}

	Sixth, due to the fixed fiber sizes for SDSS spectra, it is necessary to check whether the 
collected Type-2 AGN in different samples have intrinsically different host galaxy properties by the 
following collected parameters which could affect the results shown in Fig.~\ref{ne}. The parameters 
of petroR90\_r and petroR50\_r (as the radii containing 90\% and 50\% of the Petrosian flux in SDSS 
r-band\footnote{Parameters from different bands can lead to very similar results.} photometric image) 
are collected from the public SDSS database 'PhotoObjAll' which contains full photometric catalog 
quantities, in order to check effects of size ($R_{90}$ and $R_{50}$) of host galaxy. The parameter 
of d4000\_n (the common parameter Dn(4000) as discussed in \citealt{ka03}) are collected from the public 
SDSS database 'galSpecIndx', in order to check effects of stellar ages of host galaxy. The parameter 
mstellar\_median are collected from the public database 'stellarMassPCAWiscM11' as described in 
\citet{ms11, ch12}, in order to check effects of total stellar mass $M_{*}$ of host galaxy. 
The detailed descriptions on the applied databases above can be found in 
\url{https://skyserver.sdss.org/dr16/en/help/docs/tabledesc.aspx}. Then, Fig.~\ref{n1ew} shows the 
distributions of $R_{90}$, $R_{50}$, Dn(4000) and $M_{*}$ for the Type-2 AGN in the Sample-I and in 
the Sample-II, respectively, leading to clearly different distributions of the parameters of the Type-2 
AGN in the different samples. In order to check the effects of the different distributions of the 
parameters shown above, a new subsmaple-I including 96 Type-2 AGN from the Sample-II and a new 
subsample-II including 96 Type-2 AGN from the Sample-I can be created to have the same distributions 
of redshift $z$, $R_{90}$, $R_{50}$, Dn(4000) and $M_{*}$. Here, combining with the same distributions 
of size of host galaxy, to also consider the same redshift distributions can be applied to ignore 
aperture effects. The first five panels of Figure~\ref{n2ew} shows the same distributions of $z$, $R_{90}$, 
$R_{50}$, Dn(4000) and $M_{*}$ of the Type-2 AGN in the new subsamples. Through the two-sided 
Kolmogorov-Smirnov statistic technique, the 96 Type-2 AGN in the new subsample-II and the 96 Type-2 
AGN in the new subsample-I have the same distributions of redshift, size of host galaxy, total stellar 
mass and stellar age, with significance levels higher than 95.3\%. Here, the same distributions of 
$R_{90}$ and $R_{50}$ also indicate the similar distributions of inverse concentration parameter 
\citep{ref29, ref30} to show not large difference of host galaxy morphology of the collected Type-2 
AGN. Then, bottom right panel of Figure~\ref{n2ew} shows the $R_{S2}$ distributions of the Type-2 AGN 
in the new subsamples. The median values and the corresponding bootstrap method determined 
uncertainties of $\log(R_{S2})$ (with $R_{S2}~<~1.5$) are about $(10.31\pm0.28)\times10^{-2}$ and 
$(5.68\pm0.15)\times10^{-2}$ for the Type-2 AGN in the new subsample-II and the Type-2 AGN in the new 
subsample-I, respectively. And through the Wilcoxon Rank-Sum Test, the Type-2 AGN in the 
subsample-II and in the subsample-I have significantly different median values of $\log(R_{S2})$ 
(with $R_{S2}~<~1.5$) with significance level about $10^{-13}$. Therefore, there are few effects of 
host galaxy properties on the different $\log(R_{S2})$ for the Type-2 AGN in the Sample-II and in the 
Sample-I.

	Seventh, although we have simply accepted few effects of orientations on NLRs in Type-2 AGN 
considering the unified model of AGN, we should note that \citet{hs18} have discussed that structure 
properties of NLRs in Type-2 AGN could be shaped by obscurations. Therefore, it is necessary to check 
effects of obscurations traced by WISE mid-IR colors (W2-W4) as pointed in \citet{hs18}. Then, through 
the SDSS public database 'WISE\_allsky', the photometric magnitudes in W2 and W4 bands are collected 
for the collected Type-2 AGN. Left panel of Fig.~\ref{wise} shows the mid-IR color distributions of the 
Type-2 AGN, with the mean value of about 5.42 for the Type-2 AGN in the Sample-II and about 5.64 for the 
Type-2 AGN in the Sample-I. And through the Students t-statistic technique, the Type-2 AGN in the 
Sample-II and in the Sample-I have the same mean values of W2-W4 with significance level only about 
$2\times10^{-2}$, indicating it is necessary to check effects of different W2-W4 distributions. Among 
the Type-2 AGN in the Sample-II and in the Sample-I, two subsamples, one subsample (subsample-II) 
including 1049 Type-2 AGN from the Sample-II and the other subsample (subsample-I) including 1049 
Type-2 AGN from the Sample-I, can be simply created to have the same W2-W4 distributions with significance 
levels higher than 99.99\% through the two-sided Kolmogorov-Smirnov statistic technique. The same W2-W4 
distributions of Type-2 AGN in the new subsamples are shown in middle panel of Fig.~\ref{wise}. Meanwhile, 
right panel of Fig.~\ref{wise} shows $\log(R_{S2})$ distributions of the Type-2 AGN in the two new 
subsamples. The median values and the corresponding bootstrap method determined uncertainties 
of $\log(R_{S2})$ (with $R_{S2}<1.5$) are about $(8.99\pm0.15)\times10^{-2}$ and $(6.81\pm0.18)\times10^{-2}$ 
for the Type-2 AGN in the new subsample-II and in the new subsample-I, respectively. And 
through the Wilcoxon Rank-Sum Test, the Type-2 AGN in the new subsample-II and in the new subsample-I 
have the same median values with significance level only about $10^{-11}$, to re-confirm higher $R_{S2}$ 
for the Type-2 AGN above the $R_{NLRs}-L_{O3}$ relation. Therefore, there are few effects of obscuration 
shaped NLRs structures on the different $\log(R_{S2})$ for the Type-2 AGN in the Sample-II and in the Sample-I.

	Therefore, the results shown in Fig.~\ref{ne} and Fig.~\ref{szd} and corresponding discussions 
above can be applied to support the $R_{NLRs}-L_{O3}$ empirical relation reported in \citet{lz13} and 
also to support electron density gradients in AGN NLRs. Unfortunately, only through comparisons between 
$R_{fib}$ and $R_{NLRs}$, it is hard to estimate a clear quantitative result on the expected 
$R_{NLRs}-L_{O3}$ relation and the electron density gradients in AGN NLRs.

	Before end of the subsection, three additional points are noted. First, besides the 
discuused median values of $\log(R_{S2})$ for the AGN in the different samples/subsamples, mean values 
of $\log(R_{S2})$ have also been checked, and the Students t-statistic technique can be applied to 
confirm the AGN in the different samples/subsamples having the very different mean values of $\log(R_{S2})$ 
with confidence levels higher than 5$\sigma$. Second, besides the discussed subsamples above with the same 
distributions of the pointed parameters, similar subsamples are also created with the same distributions 
of the same parameters, but with different objects included, after randomly shuffling the orders of the 
objects in the parent samples. Then, different subsamples with the same distributions of the given 
parameters lead to the very similar $\log(R_{S2})$ difference, to support few effects of the randomly 
collected objects on our final results. Third, besides the 
$R_{NLRs}-L_{O3}^{0.25}$ relation reported in \citet{lz13}, the $R_{NLRs}-L_{O3}^{0.4\sim0.5}$ relation 
discussed in \citet{sc03, fk18, cs19} is also shown in Fig.~\ref{rl}. Unfortunately, in Fig.~\ref{rl}, 
there is only one Type-2 AGN lying below the $R_{NLRs}-L_{O3}$ relation minus 2RMS scatter reported in 
\citet{sc03, fk18}, and only four Type-2 AGN lying below the $R_{NLRs}-L_{O3}$ relation minus its 
corresponding 2RMS scatter reported in \citet{cs19}. In other words, if accepted the $R_{NLRs}-L_{O3}$ 
relation reported in \citet{sc03, fk18}, almost all the collected SDSS Type-2 AGN have their NLRs totally 
covered by SDSS fibers, indicating similar properties of electron densities in NLRs of the Type-2 AGN 
in the Sample-I and in the Sample-II created above, against the shown results in Fig.~\ref{ne}. Therefore, 
rather than the $R_{NLRs}-L_{O3}^{0.4\sim0.5}$ relation reported in \citet{sc03, fk18, cs19}, the 
$R_{NLRs}-L_{O3}^{0.25}$ relation reported in \citet{lz13} is the preferred empirical relation to estimate 
NLRs sizes of SDSS AGN through the [O~{\sc iii}] line luminosity measured from SDSS fiber optical 
spectroscopic results.

\section{Summary and Conclusions}

   The main summary and conclusions are as follows. 
\begin{itemize}   
\item Considering AGN unified model expected spatial properties of NLRs of Type-2 AGN, 
	SDSS fiber radii provided spatial distance $R_{fib}$ can be applied to test the 
	$R_{NLRs}-L$ empirical relation to estimate NLRs sizes of AGN.
\item Comparing $R_{fib}$ and $R_{NLRs}$ estimated through [O~{\sc iii}] line luminosity, 
	Type-2 AGN in SDSS can be divided into two samples, one sample of Type-2 AGN with 
	$R_{fib}>R_{NLRs}$ have their NLRs totally covered by SDSS fibers, the other sample 
	of Type-2 AGN with $R_{fib}<R_{NLRs}$ have their NLRs partly covered by SDSS fibers.
\item Considering expected electron density gradients in AGN NLRs, different physical 
	properties of NLRs could be expected for Type-2 AGN with $R_{fib}>R_{NLRs}$ and 
	for Type-2 AGN with $R_{fib}<R_{NLRs}$.
\item Among the SDSS pipeline classified Type-2 AGN, after LINERs being removed, considering 
	the $R_{NLRs}-L$ empirical relation and 2RMS scatters reported in \citet{lz13}, 
	3658 Type-2 AGN with $R_{fib}>R_{NLRs}$ are collected and included in Sample-II, 
	and 1062 Type-2 AGN $R_{fib}<R_{NLRs}$ are collected and included in Sample-I.
\item There are statistically lower $R_{S2}$ (flux ratio of [S~{\sc ii}]$\lambda6717$\AA~ 
	to [S~{\sc ii}]$\lambda6731$\AA) for the Type-2 AGN in Sample-I than in Sample-II, 
	with confidence level higher than 5$\sigma$.
\item There are no different dependence of $R_{S2}$ on redshift for the Type-2 AGN in 
	Sample-I and in Sample-II, indicating few effects of different redshift distributions 
	on statistically lower $R_{S2}$ for the Type-2 AGN in Sample-I than in Sample-II.
\item Through the Type-2 AGN in Sample-I and in Sample-II, two subsamples can be created to 
	have the same redshift distributions. And statistically lower $R_{S2}$ can be 
	confirmed for the Type-2 AGN in the subsample-I than in the subsample-II, to reconfirm 
	few effects of different redshift distributions on statistically lower $R_{S2}$ for 
	the Type-2 AGN in Sample-I than in Sample-II.
\item Through the Type-2 AGN in Sample-I and in Sample-II, two subsamples can 
	be created to have the same distributions of $z$, $R_{90}$, $R_{50}$, $M_{*}$ and 
	Dn(4000) to check effects of host galaxy properties, to reconfirm the statistically 
	lower $R_{S2}$ for the Type-2 AGN in Sample-I.
\item Even after considering effects of obscurations traced by WISE mid-UR 
	colors on shaped NLRs structures, through two subsamples with the same W2-W4 
	distributions, the statistically lower $R_{S2}$ can be reconfirmed for the Type-2 
	AGN in Sample-I. 
\item Considering NLRs partly covered by SDSS fibers for the Type-2 AGN in Sample-I, more 
	apparently statistically lower intrinsic $R_{S2}$ could be expected for the Type-2 
	AGN in Sample-I. 
\item Different dependence of $R_{S2}$ on distance $Dis$ of objects to the $R_{NLRs}-L_{O3}$ 
	empirical relation in \citet{lz13} can be found for the Type-2 AGN in Sample-I and 
	in Sample-II, providing clues to support that the provided method in this manuscript 
	can be efficiently applied to test the $R_{NLRs}-L_{O3}$ empirical relation.
\item Statistically lower $R_{S2}$ for the Type-2 AGN in Sample-I can be applied to support 
	the $R_{NLRs}-L_{O3}$ empirical relation in \citet{lz13} to estimate NLRs sizes of 
	SDSS AGN, and also to support the commonly expected electron density gradients in 
	AGN NLRs.
\end{itemize}

\section*{Acknowledgements}
Zhang gratefully acknowledge the anonymous referee for giving us constructive 
comments and suggestions to greatly improve the paper. Zhang gratefully thanks the kind 
financial support from GuangXi University and the kind grant support from NSFC-12173020 and 
NSFC-12373014. This manuscript has made use of the data from the SDSS projects. The 
SDSS-III web site is http://www.sdss3.org/. SDSS-III is managed by the Astrophysical 
Research Consortium for the Participating Institutions of the SDSS-III Collaborations. 

\section*{Data Availability}
The data underlying this article will be shared on reasonable request to the corresponding 
author (\href{mailto:xgzhang@gxu.edu.cn}{xgzhang@gxu.edu.cn}).

\label{lastpage}
\end{document}